\documentclass[twoside,11pt]{9ssmmp} 
\usepackage{fancyhdr}
\usepackage{graphicx}
\usepackage{amssymb}
\usepackage[all]{xy}
\begin{document}


\title{Aspects of braneworld cosmology and holography\hspace{.25mm}\thanks{\,Talk presented at
IX Mathematical Physics Meeting: School and Conference on Modern Mathematical Physics,
Belgrade, 18-23 September 2017}}


\author{
\bf{Neven Bili\'c}\hspace{.25mm}\thanks{\,e-mail address: bilic@irb.hr} \\
\normalsize{Division of Theoretical Physics,  Rudjer Bo\v{s}kovi\'c Institute}\\
\normalsize{P.O.\ Box 180, 10001 Zagreb, Croatia} \vspace{2mm} \\
}

\date{} 

\maketitle 

\begin{abstract}
In a holographic  braneworld universe a cosmological fluid occupies a 3+1 dimensional brane
located at  the boundary of the asymptotic  AdS$_5$ bulk. The AdS/CFT correspondence 
and the second Randall-Sundrum model are combined to establish a relationship
between the RSII braneworld cosmology  and the boundary metric induced by the time dependent bulk geometry. 
Some physically interesting scenarios are  discussed in the framework of the Friedmann Robertson Walker cosmology  
involving the RSII and holographic braneworlds.
\end{abstract}

\tableofcontents

\section{Introduction}

Branewarld cosmology is based on the  scenario in which matter is confined 
on a brane moving in the higher dimensional bulk
with only gravity allowed to propagate in the bulk  \cite{randall1,randall2,arkani,antoniadis}.
The brane can be placed, e.g., at the boundary of
a 5-dim asymptotically Anti de Sitter space (AdS$_5$).
Anti de Sitter space is dual to a conformal field theory at
its boundary through the so called AdS/CFT correspondence \cite{maldacena}.
This correspondence reflects an obvious symmetry relationship:
On the one hand, AdS$_5$ is a maximally symmetric solution to Einstein’s
equations with negative cosmological constant with 
the symmetry group  AdS$_5$ $\equiv$ SO(4,2).
On the other hand,
the 3+1 boundary conformal field theory is invariant under
conformal transformations: Poincar\'{e} + dilatations + special
conformal transformation. These transformations constitute the conformal group $\equiv$ SO(4,2).
 \begin{figure}[t]
\begin{center}
\includegraphics[width=\textwidth,trim= 0 3cm 0 0]{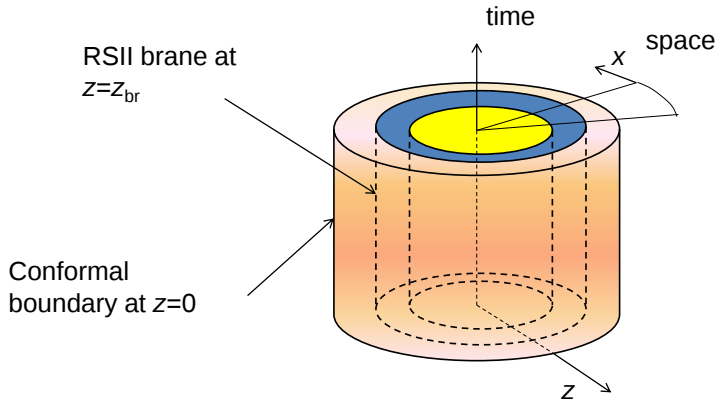}
\caption{Illustration of the AdS$_5$ bulk with two branes: RSII brane located at $z=z_{\rm br}$  
and the holographic brane at $z=0$.
}
\label{fig1}
\end{center}
\end{figure}

We will consider two types of braneworlds (Fig.~\ref{fig1}):
1) Holographic braneworld in with a 3-brane located at the boundary of the asymptotic AdS$_5$. 
The cosmology is governed by matter on the brane in addition to the boundary CFT.
2) Randall-Sundrum  braneworld with a 3-brane located  at a finite distance from the boundary of AdS$_5$. 
We will demonstrate that there exists  a map between these two substantially different scenarios.
Most of the material presented here is based on \cite{bilic1}
and earlier works \cite{kiritsis,apostolopoulos,tetradis,brax}.

We use the metric signature $(+, - - - -)$ and curvature convention
$R^{a}{}_{bcd} = \partial_c \Gamma_{db}^a - 
\partial_d \Gamma_{cb}^a + \Gamma_{db}^e \Gamma_{ce}^a  - \Gamma_{cb}^e \Gamma_{de}^a$ 
and $R_{ab} = R^s{}_{asb}$, 
so that Einstein's equations are $R_{ab} - \frac{1}{2}R G_{ab} = +8\pi G T_{ab}.$ 

\section{Randall-Sundrum model}
\subsection{Basics}
The Randall-Sundrum (RS) model \cite{randall1,randall2}.
is a simple physically relevant model related to  AdS/CFT.  
The model was originally proposed as a solution to the hierarchy problem in particle physics
and as a possible mechanism for localizing gravity on the 3+1 dimensional universe embedded in
a 4+1 spacetime without compactification of the extra dimension.
It was soon realized that the RS model
is deeply rooted in a wider framework of AdS/CFT correspondence 
\cite{gubser2,nojiri1, giddings,hawking,duff3,nojiri2,haro2}.

The Randall-Sundrum model is a 4+1-dimensional universe with AdS$_5$  
geometry containing two 3-branes with opposite brane 
tensions separated in the 5th dimension. The total action is a sum
\begin{equation}
 S=S_{\rm bulk}+S_{\rm GH}+S_{\rm br1}+S_{\rm br2} ,
 \label{eq0002}
\end{equation}
where 
\begin{equation} 
S_{\rm bulk} =\frac{1}{8\pi G_5} \int d^5x \sqrt{G} 
\left[-\frac{R^{(5)} }{2} -\Lambda _5 \right], 
\label{eq001} 
\end{equation}
is the bulk action, $\Lambda_5$ being the bulk cosmological constant related to the 
AdS curvature radius as $\Lambda_5=-6/\ell^2$.
The remaining terms are the Gibbons-Hawking boundary term 
\begin{equation} 
S_{\rm GH}[h] =\frac{1}{8\pi G_5}\int_\Sigma d^{4}x\sqrt{-h} K[h] .
\label{eq003}
\end{equation} 
and two brane actions of the form
\begin{equation} 
S_{\rm br}[h] =\int_\Sigma d^{4}x\sqrt{-h} (-\sigma + \mathcal{L}^{\rm matt}[h]).
\label{eq1005}
\end{equation} 
Here we denote by $G$ the determinant of the bulk metric $G_{\mu\nu}$, by $h$  
the determinant of the metric $h_{\mu\nu}$ induced on the hypersurface $\Sigma$, and
by $\sigma$ the brane tension. Matter on the brane is described by the Lagrangian
$\mathcal{L}^{\rm matt}$.

In the following we will make use of various  coordinate systems:
\begin{enumerate}
\item
Fefferman-Graham coordinates
 \begin{equation}
ds_{(5)}^2=G_{ab}dx^adx^b =\frac{\ell^2}{z^2}\left( g_{\mu\nu} dx^\mu dx^\nu -dz^2\right),
 \label{eq3001}
\end{equation}
\item
Gaussian normal coordinates
  \begin{equation}
ds_{(5)}^2=e^{-2y/\ell} g_{\mu\nu} dx^\mu dx^\nu -dy^2,
 \label{eq5001}
\end{equation}
\item
Schwarzschild coordinates
\begin{equation}
ds_{\rm ASch}^2= 
f(r) dt^2- \frac{dr^2}{f(r)} -r^2 d\Omega_\kappa^2, 
\label{eq3202}
\end{equation}
where 
\begin{equation}
f(r)=\frac{r^2}{\ell^2}+\kappa -\mu \frac{\ell^2}{r^2},
\label{eq3225}
\end{equation}
and
\begin{equation}
d\Omega^2_\kappa=d\chi^2+\frac{\sin^2(\sqrt{\kappa}\chi)}{\kappa}(d\vartheta^2+\sin^2 \vartheta d\varphi^2)
\label{eq1004}
\end{equation}
is the spatial line element for a 
closed ($\kappa=1$), open hyperbolic ($\kappa=-1$), or open flat ($\kappa=0$) space.
The dimensionless parameter $\mu$ is  related to the black-hole mass via \cite{myers,witten2}
\begin{equation}
\mu=\frac{8G_5 M_{\rm bh}}{3\pi \ell^2}.
 \label{eq3105}
\end{equation}
\end{enumerate}
These coordinate representations are related via simple coordinate transformations
\begin{equation}
z=e^{y/\ell},  \quad
\frac{r^2}{\ell^2}=  \frac{\ell^2}{z^2}- \frac{\kappa}{2} +\frac{\kappa^2+4\mu}{16}\frac{z^2}{\ell^2}.
\label{eq3203}
\end{equation}
\subsection{Second Randall-Sundrum model (RSII)}
The RSII model  \cite{randall2} was proposed as an alternative to compactification of extra
dimensions. 
A compactification of extra dimensions is necessary to localize
gravity on the 3+1 dimensional universe.
If extra dimensions were large that would yield
unobserved modification of Newton's gravitational law. Experimental
bound on the volume of $n$ extra dimensions is \cite{long}
\begin{equation}
V^{1/n} \leq 0.1 {\rm mm}.
 \label{eq5002}
\end{equation}
RSII brane-world does not rely on compactification to localize gravity at
the brane, but on the curvature of the bulk (``warped compactification”).
The negative cosmological constant $\Lambda_5$ acts to “squeeze” the
gravitational field closer to the brane. One can see this in Gaussian
normal coordinates (\ref{eq5001}) with an exponentially attenuating warp factor $e^{-2\ell y}$.

In RSII observers reside on the positive tension brane at
$y=0$ and the negative tension brane is pushed off to
infinity in the fifth dimension.
In the original  RSII model one assumes
the $Z_2$ symmetry  $z\leftrightarrow z_{\rm br}^2/z$,  so
the region $0<z\leq z_{\rm br}$ is identified with $z_{\rm br} \leq z <\infty$, 
with the observer brane at the fixed point $z=z_{\rm br}$.
Hence, the braneworld is sitting between two patches of AdS$_5$, one on either side, and is
therefore dubbed ``two-sided'' \cite{duff3,haro2}.
In contrast, in the ``one-sided'' 
 RSII model the region $0\leq z\leq z_{\rm br}$ is simply cut off 
so the bulk is the section of spacetime $z_{\rm br} \leq z <\infty$.

The Planck mass scale is determined by the curvature of
the five-dimensional space-time
\begin{equation} 
\frac{1}{G_{\rm N}}= \frac{\gamma}{G_5}\int^\infty_{y_{\rm br}}  dy \psi^2=\frac{\gamma\ell}{2G_5}.
\label{eq0014} 
\end{equation}
where we have introduced the ``sidedness" parameter $\gamma$ \cite{bilic1}
to facilitate a joint description
of the two versions of RSII model: 
the one-sided ($\gamma=1$) and two-sided ($\gamma=2$).
One usually imposes a fine tuning condition on the brane tension
\begin{equation} 
\sigma=\sigma_0\equiv \frac{3\gamma}{8\pi G_5\ell}=\frac{3}{4\pi G_{\rm N}\ell^2}.
\label{eq0012} 
\end{equation}
which eliminates the 4-dim cosmological constant.
Note that the RSII fine tuning condition does not depend on the sidedness $\gamma$ if $\sigma_0$ is
expressed in terms of the four-dimensional Newton constant.

Table top measurements of the Newton gravitational law impose a bound
on the AdS$_5$ curvature radius.
The classical 3+1 dimensional  gravity is altered on the RSII brane due to the extra dimension.
For $r\gg \ell$ 
the weak gravitational potential  created by an isolated matter source on the brane 
is given by \cite{garriga} 
\begin{equation} 
\Phi(r)=\frac{G_{\rm N} M}{r}\left( 1+ \frac{2\ell^2}{3r^2} \right) .
\label{eq0018} 
\end{equation}
Table-top tests of Long et al \cite{long}  find no deviations of Newton's potential
at distances greater than 0.1 mm
and place the limit 
curvature
\begin{equation} 
\ell < 0.1 {\rm mm}, \quad {\rm or} \quad\ell^{-1} >10^{-12} {\rm GeV}.
\label{eq0021} 
\end{equation}

\subsection{RSII cosmology -- Dynamical brane}

Branewarld cosmology is based on the  scenario in which matter is confined 
on a brane moving in the higher dimensional bulk
with only gravity allowed to propagate in the bulk  \cite{randall1,randall2,arkani,antoniadis}.
In this section we give a simple derivation of the RSII braneworld cosmology
following  J.~Soda \cite{soda}.
Cosmology on the brane is obtained by allowing the brane to
move in the bulk. Equivalently, one could keep the brane fixed at $y=0$
while making the metric in the bulk time dependent.

Consider a time dependent brane hypersurface defined by
\begin{equation}
r-a(t)=0,
\label{eq012}
\end{equation}
in AdS-Schwarzschild background \cite{gomez,birmingham} where $a=a(t)$ is an arbitrary positive function. 
The induced line element
on the brane is
\begin{equation}
ds_{\rm ind}^2=n^2(t)dt^2 -a(t)^2 d\Omega_\kappa^2 ,
\label{eq015}
\end{equation}
where
\begin{equation}
n^2 =f(a)-\frac{(\partial_t a)^2}{f(a)},
\label{eq026}
\end{equation}
and $f$ is defined by (\ref{eq3225}).
The junction conditions on the brane with matter
\begin{equation}
 \left. K_{\mu\nu}\right|_{r=a-\epsilon}=\frac{8 \pi G_5}{3\gamma}  (\sigma g_{\mu\nu}+ 3T_{\mu\nu}) 
\label{eq018}
\end{equation}
yield
\begin{equation}
\frac{(\partial_t a)^2}{n^2 a^2} +\frac{f}{a^2}=\frac{1}{\ell^2 \sigma_0^2}(\sigma+\rho)^2 .
\label{eq020}
\end{equation}
Now, imposing the fine tuning condition (\ref{eq0012}) 
one finds a modified Friedmann equation \cite{binetruy,kraus,flanagan,mukohyama}.
\begin{equation}
\mathcal{H}^2 =\frac{8\pi G_{\rm N}}{3}\rho+\left( \frac{4\pi G_{\rm N}\ell}{3}\right)^2 \rho^2 
+\frac{\mu\ell^2}{a^4} ,
\label{eq029}
\end{equation}
where 
\begin{equation}
 \mathcal{H}^2=H^2+\frac{\kappa}{a^2}, \quad
 H=\frac{\partial_t a}{n a}.
\label{eq023}
\end{equation}
Equation (\ref{eq029}) differs from the standard  Friedmann equation by the last two terms on the right-hand side.
RSII cosmology is thus subject to astrophysical and cosmological tests
(see, e.g., Refs. \cite{maartens,godlowski}).
The deviation proportional to $\rho^2$  poses no  problem
as it decays as $a^{-8}$ in
the radiation epoch and  will rapidly become negligible after the end of the
high-energy regime $\rho\simeq \sigma_0$.
The last term on the right-hand side, the so called ``dark radiation'', for positive $\mu$
 should not
exceed 10\% of the total radiation content in
the epoch of BB nucleosynthesis  whereas for negative $\mu$ could be as large as
the rest of the radiation content \cite{ichiki,bratt}.
As expected, both one-sided and two-sided versions of the RSII model yield identical braneworld cosmologies.

The second Friedmann equation may be easily obtained by combining the time derivative of (\ref{eq029})
with the energy conservation equation
\begin{equation}
\partial_t\rho+3(\rho+p)\frac{\partial_t a}{a}=0.
 \label{3201}
\end{equation}

\section{Connection with AdS/CFT}
AdS/CFT correspondence is a holographic duality between
gravity in d+1-dimensional space-time and quantum conformal field theory (CFT) on the d-dim
boundary. Original formulation stems from string theory:
the original AdS/CFT conjecture establishes 
 an equivalence of a four dimensional ${\mathcal{N}}=4$ supersymmetric Yang-Mills theory
and string theory in a ten dimensional ${\rm AdS}_5\times {\rm S}_5$
bulk \cite{maldacena,gubser,witten1}.

\subsection{RSII braneworld as a cutoff in AdS$_5$}
In the RSII model by introducing the boundary in AdS$_5$ at
$z = z_{\rm br}$ instead of $z = 0$, the model is conjectured to be dual to
a cutoff CFT coupled to gravity, with $z = z_{\rm br}$ providing the IR
cutoff (corresponding to the UV cutoff of the boundary CFT) \cite{duff3}.
In the one-sided RSII model,  the model involves a single CFT at the
boundary of a single patch of AdS$_5$.
In the two-sided RSII model one would instead require two copies
of the CFT, one for each of the AdS$_5$ patches.

The on-shell bulk action
\begin{equation} 
S_{\rm bulk} =\frac{1}{8\pi G_5} \int d^5x \sqrt{G} 
\left[-\frac{R^{(5)} }{2} -\Lambda _5 \right], 
\label{eq0001} 
\end{equation}
is infrared divergent because physical distances diverge at $z=0$.
The asymptotically AdS metric near $z=0$ can be expanded as
\begin{equation}
ds_{(5)}^2=\frac{\ell^2}{z^2}\left( g_{\mu\nu} dx^\mu dx^\nu -dz^2\right),
 \label{eq4011}
\end{equation}
\begin{equation}
 g_{\mu\nu}=g^{(0)}_{\mu\nu}+z^2 g^{(2)}_{\mu\nu}+z^4 g^{(4)}_{\mu\nu}+z^6 
g^{(6)}_{\mu\nu}+\ldots \, .
\label{eq3002}
\end{equation}
Explicit expressions for $g^{(2n)}_{\mu\nu}$, $n=1,2,3$ 
in terms of arbitrary $g^{(0)}_{\mu\nu}$
may be found  in Ref.\ \cite{haro}.
In particular,  we will need 
 \begin{equation}
g^{(2)}_{\mu\nu}=\frac12 \left(R_{\mu\nu}-\frac16 R g^{(0)}_{\mu\nu}\right)
\label{eq3121}
\end{equation}
 and the relation
\begin{equation}
{\rm Tr} g^{(4)}=-\frac14  
{\rm Tr} (g^{(2)})^2,  
\label{eq3120}
\end{equation}
where the trace of a tensor $A_{\mu\nu}$ is defined as
\begin{equation}
{\rm Tr} A= A_\mu^\mu=
g^{(0)\mu\nu}A_{\mu\nu}  . 
\label{eq3129}
\end{equation}

Now, we regularize the action by placing the RSII brane near
the AdS boundary, i.e., at $z = \epsilon \ell$, $\epsilon\ll 1$, so that the induced
metric is
\begin{equation}
 h_{\mu\nu}=\frac{1}{\epsilon^2}(g^{(0)}_{\mu\nu}+\epsilon^2 \ell^2 g^{(2)}_{\mu\nu}+ \dots) .
\label{eq3003}
\end{equation}
The bulk splits in two regions: $0 \leq z <\epsilon\ell$, and $\epsilon\ell\leq z \leq \infty$. We
can either discard the region $0 \leq z <\epsilon\ell$ (one-sided
regularization, $\gamma=1$) or invoke the $Z_2$ symmetry and
identify two regions (two-sided regularization, $\gamma=2$).
Then, the regularized bulk action is
\begin{equation} 
S_{\rm bulk}^{\rm reg} =\gamma S_0^{\rm reg} =\frac{\gamma}{8\pi G_5} \int_{z\geq \epsilon\ell} d^5x \sqrt{G} 
\left[-\frac{R^{(5)} }{2} -\Lambda _{\left(5\right)} \right] 
+S_{\rm GH}[h]
\label{eq1006} 
\end{equation}
The renormalized action is obtained by adding
counterterms to $S_0^{\rm reg}$  and  taking the limit $\epsilon\rightarrow 0$ \cite{haro,hawking2}
\begin{equation} 
S_0^{\rm ren}[g^{(0)}]=\lim_{\epsilon\rightarrow 0} (S_0^{\rm reg}[G]+S_1[h]+S_2[h]+S_3[h]),
\label{eq1007} 
\end{equation}
The necessary counterterms are \cite{haro}
\begin{equation} 
S_1[h]=-\frac{6}{16\pi G_5\ell}\int d^4x \sqrt{-h} , 
\label{eq4001} 
\end{equation}
\begin{equation} 
S_2[h]=-\frac{\ell}{16\pi G_5}\int d^4x \sqrt{-h}\left(-\frac{R[h]}{2} \right) ,
\label{eq4002} 
\end{equation}
\begin{equation} 
S_3[h]=-\frac{\ell^3}{16\pi G_5}\int d^4x \sqrt{-h}\frac{\log\epsilon}{4}
 \left( R^{\mu\nu}[h]R_{\mu\nu}[h] -\frac13 R^2[h] \right).
\label{eq4003} 
\end{equation}
Now we demand that the variation with respect to the induced metric $h_{\mu\nu}$ of 
the total RSII action  (the sum of the
regularized on shell bulk action and the brane action (\ref{eq1005}))  vanishes, i.e., we require
\begin{equation} 
\delta ( S_{\rm bulk}^{\rm reg}[h]+S_{\rm br}[h])=0,
\label{eq4004} 
\end{equation}
which may be expressed as
\begin{eqnarray}
\delta\left[\gamma S_0^{\rm ren}- \gamma S_3-\left(\sigma-
\frac{3\gamma}{8\pi \ell G_5}\right)\int d^4x \sqrt{-h}
 +\int d^4x \sqrt{-h}\mathcal{L}_{\rm matt}\right.
\nonumber\\
\left.
+\frac{\gamma \ell}{16\pi G_5}\int d^4x \sqrt{-h}\frac{R[h]}{2} \right]=0.
\label{eq4005} 
\end{eqnarray}
The third term gives the contribution to the cosmological constant and may be eliminated 
by imposing the RSII fine tuning condition (\ref{eq0012}).
The variation  of the scheme dependent $S_3$ may be combined with the first term.
Then, according to the AdS/CFT prescription 
by functionally differentiating the renormalized on-shell bulk gravitational action with respect to the
boundary metric $g^{(0)}_{\mu\nu}$ one obtains the expectation value $\langle T^{\rm CFT}_{\mu\nu}\rangle$
and hence
\begin{equation}
\delta (S_0^{\rm ren}-S_3)= \frac12 \int d^4x \sqrt{-h} 
\langle T^{\rm CFT}_{\mu\nu}\rangle \delta {h}^{\mu\nu} ,
\label{eq4006}
\end{equation}
With this
the variation of the action yields Einstein’s equations on the
boundary
\begin{equation}
R_{\mu\nu}- \frac12 R g_{\mu\nu}= 8\pi G_{\rm N} (\gamma \langle T^{\rm CFT}_{\mu\nu}\rangle +T^{\rm matt}_{\mu\nu}),
 \label{eq3006}
\end{equation}
where the energy-momentum tensor $T^{\rm matt}_{\mu\nu}$ describes matter on the holographic brane 
in addition to the holographic conformal part given by \cite{haro}
\begin{eqnarray}
 \langle T^{\rm CFT}_{\mu\nu}\rangle = -\frac{\ell^3}{4\pi G_5}\left\{
 g^{(4)}_{\mu\nu}-\frac18 \left[({\rm Tr} g^{(2)})^2-{\rm Tr} (g^{(2)})^2\right]g^{(0)}_{\mu\nu}
 \right.
\nonumber\\
\left.
 -\frac12 (g^{(2)})^2_{\mu\nu}+\frac14 {\rm Tr} g^{(2)}g^{(2)}_{\mu\nu}
  \right\}.
 \label{eq3106}
\end{eqnarray}
This is an explicit realization of the AdS/CFT
correspondence: the vacuum expectation value of a boundary CFT operator is
obtained solely in terms of geometrical quantities of the bulk.

\subsection{Conformal anomaly}
It is of particular interest to check whether the stress
tensor $T^{\rm CFT}$ obtained using  
AdS/CFT prescription correctly reproduces the conformal anomaly. 
From (\ref{eq3106})
with the help of (\ref{eq3121}) and (\ref{eq3120}) we find
\begin{equation}
\langle {T^{\rm CFT}}^\mu_\mu\rangle =\frac{\ell^3}{128\pi G_5}
 \left( G_{\rm GB}-C^2\right) ,
\label{eq3123}
\end{equation}
where
\begin{equation}
G_{\rm GB}= R^{\mu\nu\rho\sigma}R_{\mu\nu\rho\sigma}-
 4 R^{\mu\nu}R_{\mu\nu} + R^2  
\label{eq3124}
\end{equation}
is the Gauss-Bonnet invariant and
\begin{equation}
C^2\equiv C^{\mu\nu\rho\sigma}C_{\mu\nu\rho\sigma}= R^{\mu\nu\rho\sigma}R_{\mu\nu\rho\sigma}
-2 R^{\mu\nu}R_{\mu\nu} + \frac13  R^2   
\label{eq3125}
\end{equation}
is the square of the Weyl tensor $C_{\mu\nu\rho\sigma}$.
This result should be compared with the
standard conformal anomaly calculated in field theory \cite{duff2}
\begin{equation}
\langle {T^{\rm CFT}}^\mu_\mu\rangle =
 b G_{\rm GB}-c C^2+ b'\Box R   .  
\label{eq3126}
\end{equation}
The two results agree if we ignore the last term
in (\ref{eq3126})
and identify
\begin{equation}
b=c=\frac{ G_5}{128\pi\ell^3}. 
\label{eq3127}
\end{equation}
The standard field theory calculations give \cite{duff2,duff}
\begin{equation}
  b=\frac{n_{\rm s}+(11/2)n_{\rm f}+62 n_{\rm v}}{360 (4\pi)^2},
\quad
  c=\frac{n_{\rm s}+3n_{\rm f}+12 n_{\rm v}}{120 (4\pi)^2}.
 \label{eq3128}
\end{equation}
where $n_{\rm s}$,  $n_{\rm f}$, $n_{\rm v}$ are the numbers of massless scalar bosons, Weyl fermions and  vector bosons,
respectively. Hence, generally $b\neq c$. However, 
in the ${\cal{N}}=4$ U($N$) super-Yang-Mills theory, 
$n_{\rm s}=6N^2$, $n_{\rm f}=4N^2$, and $n_{\rm v}=N^2$, 
in which case the equality $b=c$ holds
and the conformal anomaly is correctly reproduced by the
holographic expression (\ref{eq3123}) if we identify \cite{henningson} 
\begin{equation}
  \frac{\ell^3}{G_5}=\frac{2N^2}{\pi}.
 \label{eq3118}
\end{equation}

\section{Holographic cosmology}

As we have shown in the previous section, in the limit in which the RSII brane
approaches the AdS$_5$ boundary, the geometry on the boundary brane, 
referred to as the {\em holographic brane},
satisfies a special form of Einstein's equations (\ref{eq3006}).
To derive the corresponding cosmology  we start from AdS-Schwarzschild static coordinates 
in the bulk and
make the coordinate transformation 
\begin{equation}
 t=t(\tau,z), \quad  r=r(\tau,z).
 \label{eq204}
\end{equation}
The line element will take a general form
\begin{equation}
ds_{(5)}^2=\frac{\ell^2}{z^2}\left( 
n^2(\tau,z)d\tau^2- a^2(\tau,z) d\Omega_\kappa^2-dz^2 
\right) ,
 \label{eq102}
\end{equation}
Imposing the boundary conditions at $z=0$:
\begin{equation}
 n(\tau,0)=1, \quad a(\tau,0)= a_{\rm h}(\tau),
\end{equation}
we obtain the induced metric at the boundary in the general FRW
form
\begin{equation}
ds_{(0)}^2=g^{(0)}_{\mu\nu}dx^\mu dx^\nu =d\tau^2 -a_{\rm h}^2(\tau) d\Omega_\kappa^2 .
 \label{eq3201}
\end{equation}
Solving Einstein’s equations in the bulk one finds \cite{apostolopoulos}
\begin{equation}
a^2=a_{\rm h}^2\left[
\left(1-\frac{\mathcal{H}_{\rm h}^2 z^2}{4}\right)^2
+ \frac14 \frac{\mu z^4}{a_{\rm h}^4}
\right],
\quad
{\mathcal{N}}=\frac{\dot{a}}{\dot{a}_{\rm h}}.
 \label{eq3103}
\end{equation}
where 
\begin{equation}
 \mathcal{H}_{\rm h}^2=H_{\rm h}^2+\frac{\kappa}{a_{\rm h}^2},
 \label{eq3111}
\end{equation}
and  $H_{\rm h}=\dot{a}_{\rm h}/a_{\rm h}$ is the Hubble expansion rate  on  boundary.
Comparing the exact  $g_{\mu\nu}(\tau,z)$ in (\ref{eq102})  with the expansion (\ref{eq4011})
we can extract $g_{\mu\nu}^{(2)}$ and $g_{\mu\nu}^{(4)}$.
Then, using the 
expression (\ref{eq3106}) we obtain
\begin{equation}
 \langle T^{\rm CFT}_{\mu\nu}\rangle = t_{\mu\nu}+
\frac14 \langle {T^{\rm CFT}}^\alpha_\alpha\rangle g^{(0)}_{\mu\nu} ,
 \label{eq3107}
\end{equation}
where the second term on the right-hand side corresponds to the conformal anomaly
\begin{equation}
 \langle {T^{\rm CFT}}^\alpha_\alpha\rangle = 
\frac{3\ell^3}{16\pi G_5}\frac{\ddot{a}_{\rm h}}{a_{\rm h}}\mathcal{H}_{\rm h}^2 ,
 \label{eq3027}
\end{equation}
and the first term is a traceless tensor with non-zero components
\begin{equation}
 t_{00}=-3 t^i_i =\frac{3\ell^3}{64\pi G_5 }
\left(\mathcal{H}_{\rm h}^4 +\frac{4\mu}{a_{\rm h}^4}
-\frac{\ddot{a}_{\rm h}}{\dot{a}_{\rm h}}\mathcal{H}_{\rm h}^2\right).
 \label{eq3108}
\end{equation}
Hence, apart from the conformal anomaly, the CFT dual to
the time dependent asymptotically AdS$_5$ bulk metric is a
conformal fluid with the equation of state
$p_{\rm CFT}=\rho_{\rm CFT}/3$,
where $\rho_{\rm CFT}=t_{00}$, $p_{\rm CFT}=-t^i_i$.

Using this,
from the boundary Einstein equations we obtain the
holographic Friedmann equation \cite{apostolopoulos,kiritsis}
\begin{equation}
 \mathcal{H}_{\rm h}^2=\frac{\ell^2}{4}
 \left(\mathcal{H}_{\rm h}^4+ \frac{4\mu}{a_{\rm h}^4} \right)+\frac{8\pi G_{\rm N}}{3}\rho_{\rm h}.
 \label{eq3110}
\end{equation}
Here we have used the  energy-momentum tensor with nonvanishing components
\begin{equation}
  T^{\rm matt}_{00}= \rho_{\rm h} , \quad T^{\rm matt}_{ij}= p_{\rm h} g^{(0)}_{ij} ,
 \label{eq3109}
\end{equation}
where $\rho_{\rm h}$ and $p_{\rm h}$ are  the matter energy density and pressure,  respectively. 
The second Friedmann equation can be derived by combining the time derivative of (\ref{eq3110}) with 
the energy conservation equation
\begin{equation}
\dot{\rho}_{\rm h}+3(\rho_{\rm h}+p_{\rm h})H_{\rm h}=0. 
\end{equation}
One finds
\begin{equation}
 \frac{\ddot{a}_{\rm h}}{a_{\rm h}}
 \left(1-\frac{\ell^2}{2}\mathcal{H}_{\rm h}^2\right)+\mathcal{H}_{\rm h}^2=
 \frac{4\pi G_{\rm N}}{3}(\rho_{\rm h}-3p_{\rm h}).
 \label{eq3113}
\end{equation}

\section{Holographic map}

The time dependent bulk spacetime with metric (\ref{eq102})
may be regarded as a $z$-foliation of the bulk with FRW cosmology on each $z$-slice
\cite{bilic1}.
In particular, 
at $z=z_{\rm br}$ one has the RSII  cosmology and at  at $z=0$ the holographic cosmology. 
A map between a $z$-cosmology and $z=0$-cosmology can be constructed using
(\ref{eq3103}) and the inverse relation 
\begin{equation}
a_{\rm h}^2=\frac{a^2}{2}\left(
1+\frac{\mathcal{H}^2 z^2}{2}
+\mathcal{E}
\sqrt{1+\mathcal{H}^2 z^2- \frac{\mu z^4}{a^4}}
\right),
 \label{eq4103}
\end{equation}
where
\begin{equation}
\mathcal{E}=\left\{ \begin{array}{ll}
-1,& \mbox{for two-sided version},\\
\pm 1, & \mbox{for one-sided version} . \end{array} \right.
\label{eq4105}
\end{equation}

A functional relationship between Hubble rates can be obtained
by making use of (\ref{eq3103}) and (\ref{eq3111}). One finds
\begin{equation}
\mathcal{H}^2= \mathcal{H}_{\rm h}^2\left[1-\frac{\mathcal{H}_{\rm h}^2 z^2}{2}
+ \frac{1}{16}\left(\mathcal{H}_{\rm h}^4+\frac{4\mu}{a_{\rm h}^4}\right)z^4 
\right]^{-1}.
 \label{eq3211}
\end{equation}
The map is schematically illustrated as
$$
\xymatrix{
d\tau^2 -a_{\rm h}^2 d\Omega_\kappa^2  \ar[rr]^{\tau\rightarrow\tilde{\tau}} \ar[d]_z & & 
(1/n^2)d\tilde{\tau}^2 -a_{\rm h}^2 d\Omega_\kappa^2 \ar[d]^z \\
n^2 d\tau^2 -a^2 d\Omega_\kappa^2 \ar[rr]_{\tau\rightarrow\tilde{\tau}} & & 
d\tilde{\tau}^2 -a^2 d\Omega_\kappa^2
}
$$
where  $\tau$ and $\tilde{\tau}$ are the holographic and RSII synchronous times, respectively.

\begin{figure}[ht]
\begin{center}
\includegraphics[width=0.45\textwidth,trim= 0 0 0 0]{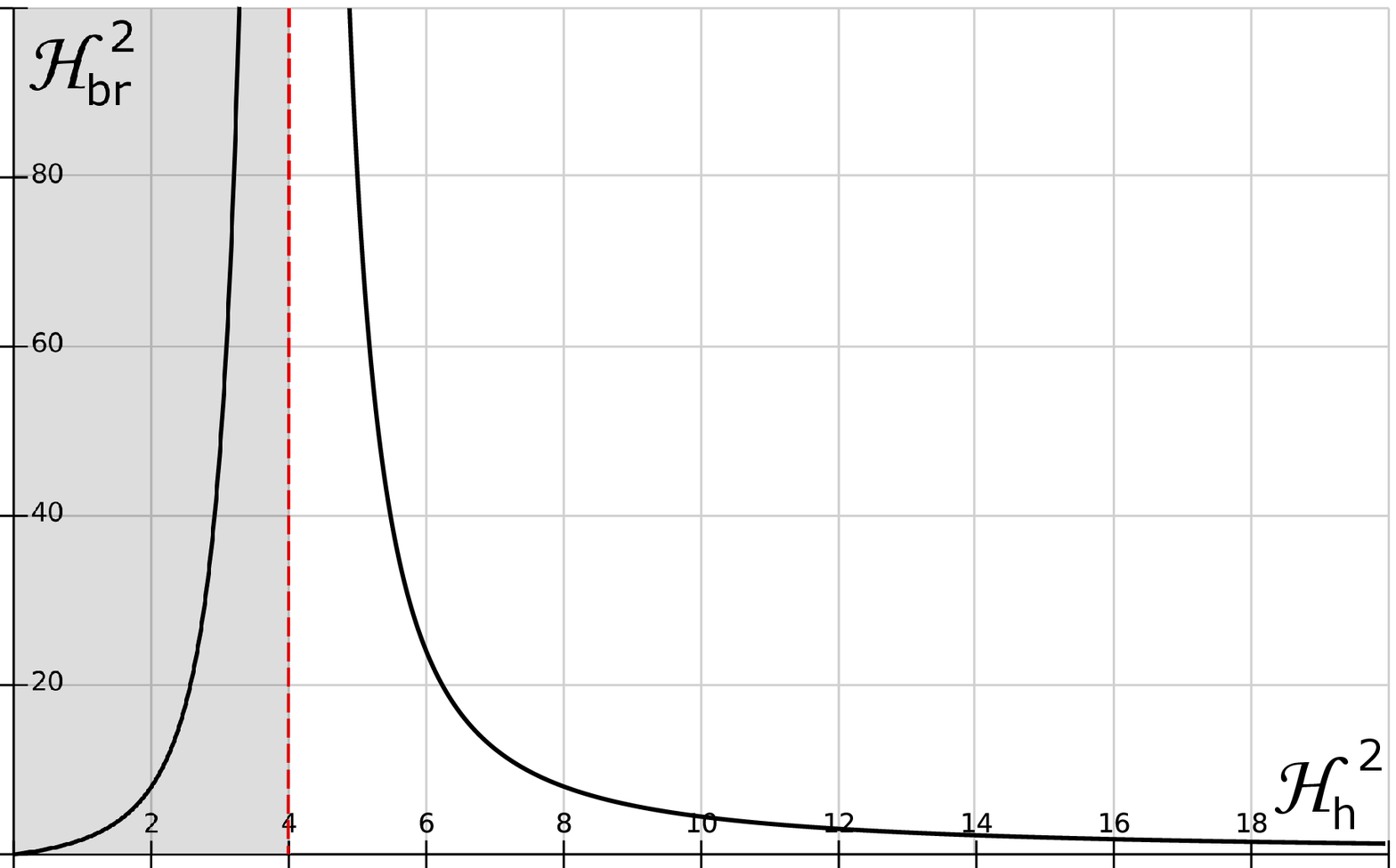}
\includegraphics[width=0.45\textwidth,trim= 0 0 0 0]{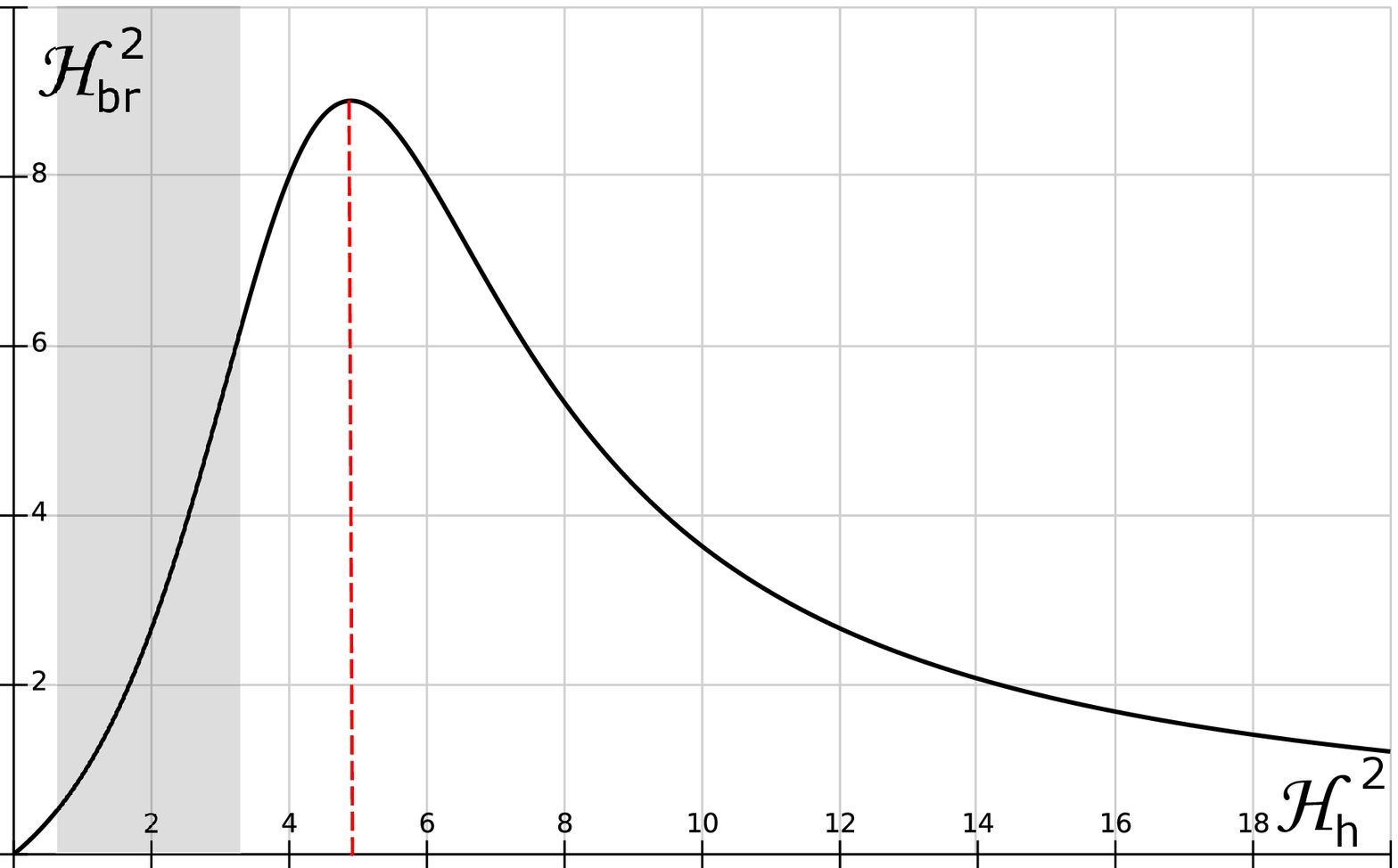}
\caption{$\mathcal{H}_{\rm br}^2$ as a function of $\mathcal{H}_{\rm h}^2$
(both in units of $z_{\rm br}^{-2}$)  defined by (\ref{eq3211}) for $\mu= 0$ (left panel)
and  $\mu \ell^4/a_{\rm h}^4=2$ with $z_{\rm br}^2/\ell^2=2$ (right panel).
The region left from the vertical dashed red line is relevant for the one-sided version only.
The shaded area corresponds to the physical region $\rho_{\rm h} >0$.
}
\label{fig2}
\end{center}
\end{figure}

As an example,  consider the RSII braneworld at $z_{\rm br}=\sqrt2 \ell$ 
and the holographic braneworld at $z=0$ with the corresponding Hubble rates $\mathcal{H}_{\rm br}^2$ 
and $\mathcal{H}_{\rm h}^2$. 
In Fig.~\ref{fig2} we plot $\mathcal{H}_{\rm br}^2$ versus
$\mathcal{H}_{\rm h}^2$ for two values  of the black hole mass parameter 
$\mu=0$ (left panel) and $\mu \ell^4/a_{\rm h}^4=1/2$ with $z_{\rm br}^2/\ell^4=2$ (right panel). 
In both panels the shaded area, corresponding to the physical region $\rho_{\rm h} >0$, 
 is determined by the condition
 \begin{equation}
2-2\sqrt{1-\mu\ell^4/a_{\rm h}^4}\leq \mathcal{H}_{\rm h}^2\ell^2\leq 2+2\sqrt{1-\mu\ell^4/a_{\rm h}^4}  .
 \label{eq4404}
\end{equation}

\section{Effective energy density}
Next, 
we analyze a few special cases in two scenarios.

\begin{enumerate}
\item
The RSII scenario with
the primary braneworld  at $z=z_{\rm br}$. 
 \item 
The holographic scenario with 
the primary cosmology  on the AdS
boundary at $z = 0$. 
\end{enumerate}
In each of the two scenarios we assume the presence of matter on the primary brane only and  no matter in the bulk.

\subsection*{RSII scenario}
In the RSII scenario the primary braneworld is the RSII brane at $z = z_{\rm br}$.
The cosmology on the $z = 0$ brane emerges as a
reflection of the RSII cosmology.
For simplicity we take $z_{\rm br}=\ell$ and we fine tune the tension $\sigma$ 
as in (\ref{eq0012}).
Then, assuming the modified Friedmann equations (\ref{eq3110}) and (\ref{eq3113})  hold on the holographic
brane, the effective energy density is given by
\begin{equation}
\frac{\rho_{\rm h}}{\sigma_0}= \frac{4\mathcal{E}(\rho/\sigma_0+1-\mathcal{E})
}{(\rho/\sigma_0+1+\mathcal{E})^2+\mu \ell^4/a^4} .
 \label{eq3217}
\end{equation}
where  $\mathcal{E}$ is defined by (\ref{eq4105}).
Given the equation of state  $p=p(\rho)$ on the RSII brane, the
cosmological scale $a$ is derived by integrating (\ref{eq029}) and (\ref{3201}).

Thus, the two-sided model with positive energy density and positive $\mu$
maps into a holographic cosmology with negative effective energy density $\rho_{\rm h}$.
For $\mu=0$ the density$\rho_{\rm h}$ diverges with $\rho$ as $1/\rho$.
The one-sided model maps into two branches: $\mathcal{E} = -1$ branch identical with
the two-sided map and the $\mathcal{E} = +1$ branch with a smooth positive function
$\rho_{\rm h}=\rho_{\rm h}(\rho)$.

\subsection*{Holographic scenario}
\begin{figure*}[ht]
\begin{center}
\includegraphics[width=0.45\textwidth ]{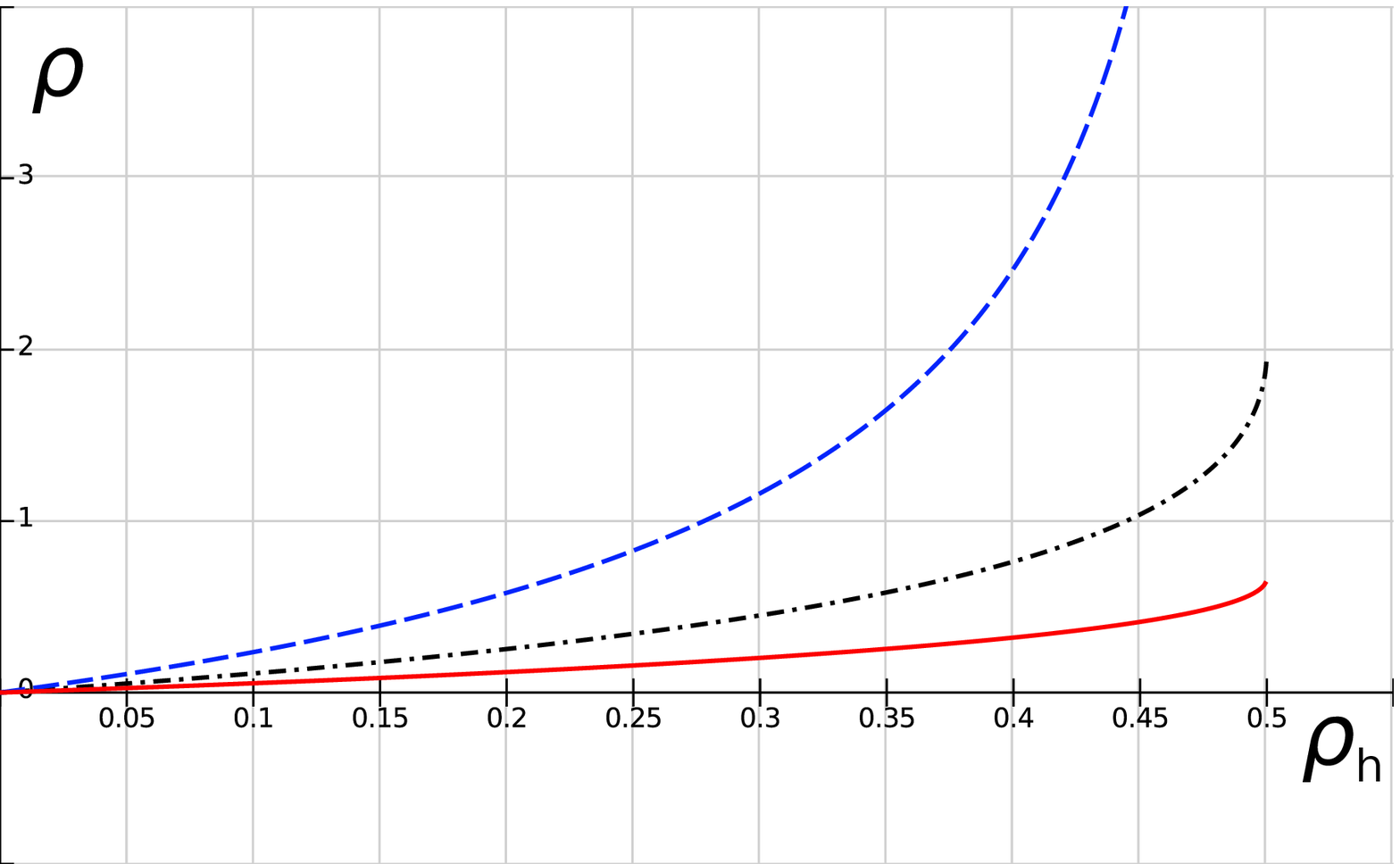}
\hspace{0.02\textwidth}
\includegraphics[width=0.45\textwidth ]{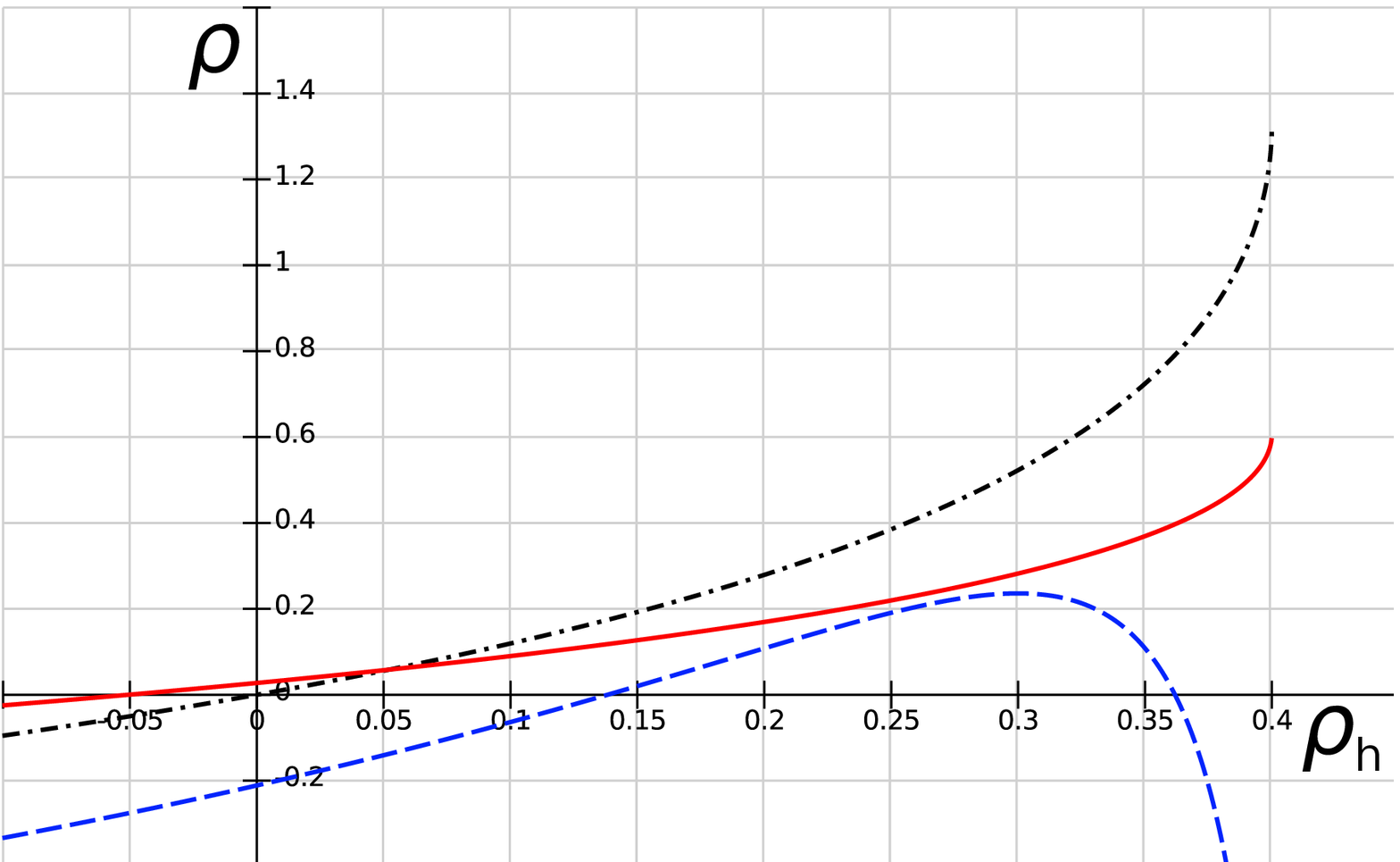}
\\
\vspace{0.02\textwidth}
\includegraphics[width=0.45\textwidth ]{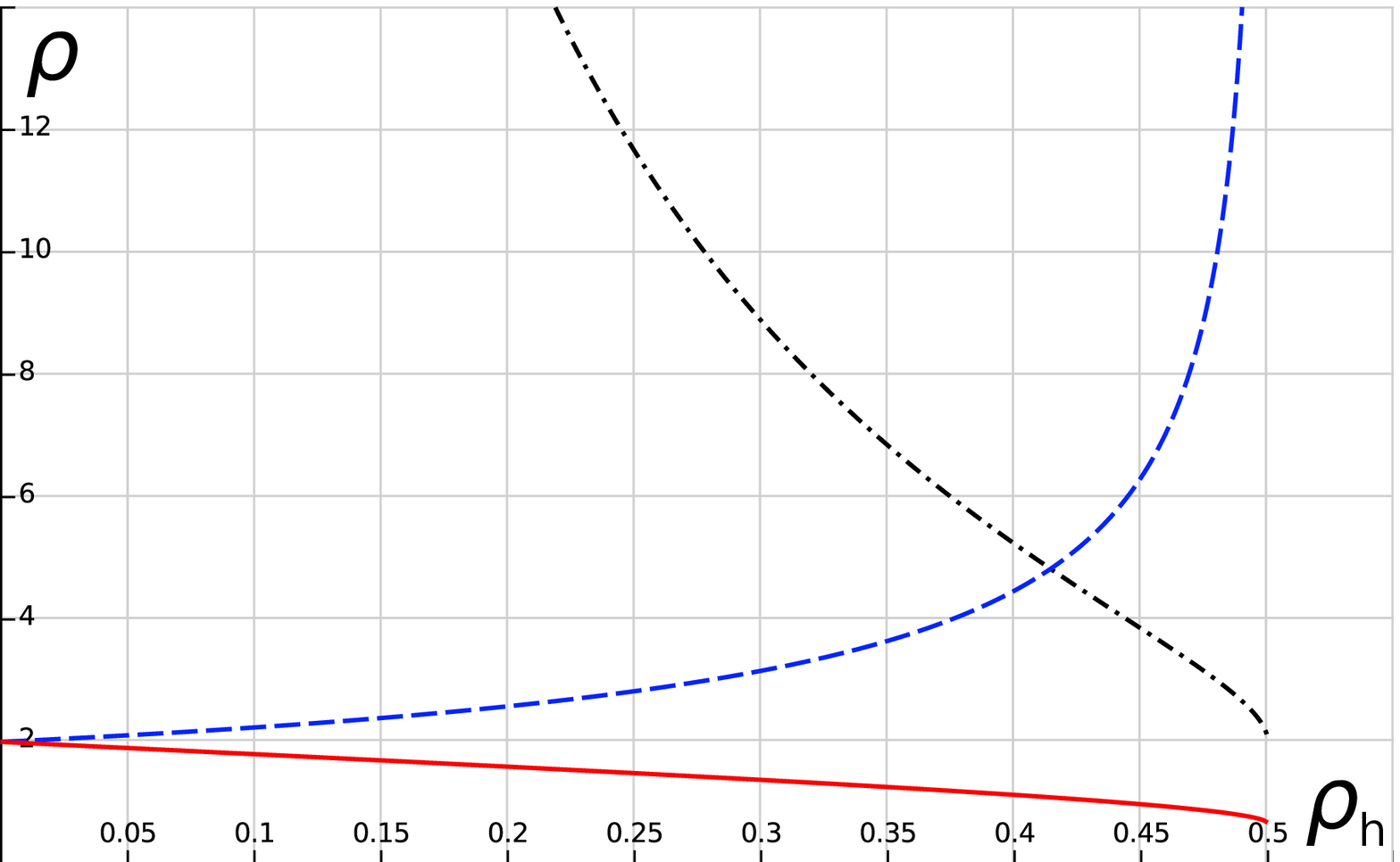}
\hspace{0.02\textwidth}
\includegraphics[width=0.45\textwidth ]{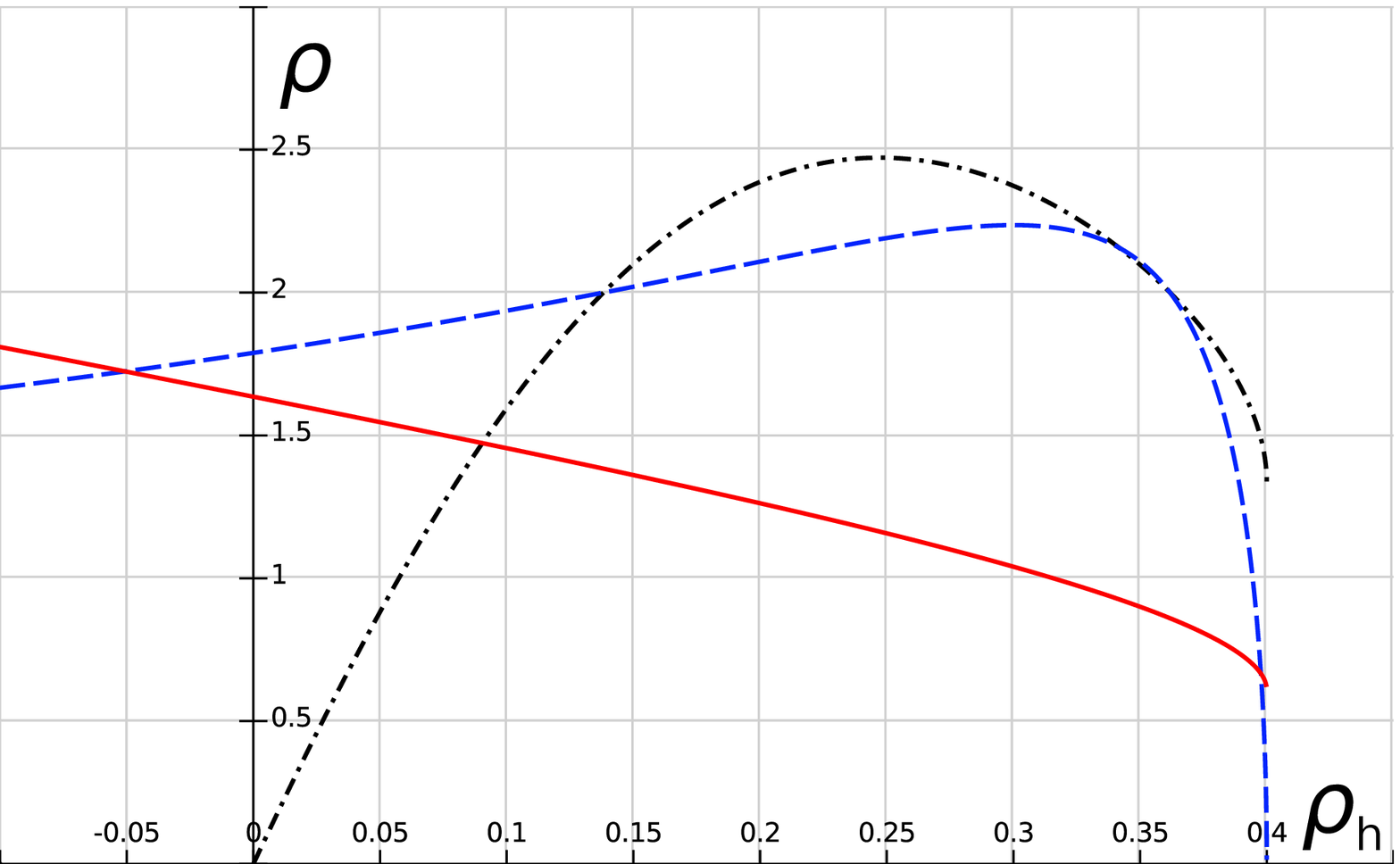}
\caption{The effective density $\rho$ on the RSII brane  as a function of the density $\rho_{\rm h}$ 
on the holographic brane 
(both in units of $\sigma_0$) for  $\sigma=\sigma_0$, $\epsilon=-1$ (top panels), $\epsilon=+1$ (bottom panels),
and $\mu \ell^4/a_{\rm h}^4=$ 0 (left panels), 0.2 (right panels).
The full red, dash-dotted black, and dashed blue lines represent $z_{\rm br}^2/\ell^2=$ 0.5, 1,  and 2, respectively.
}
\label{fig3}
\end{center}
\end{figure*}

Suppose the cosmology on the $z=0$ brane is known, i.e., the density
$\rho_{\rm h}$, the pressure $p_{\rm h}$, and the cosmological scale $a_{\rm h}$ are known. 
If there is no matter in the
bulk the induced cosmology on an arbitrary $z$-slice will be completely
determined. 
Observers on the RSII brane on an
arbitrary $z$-slice experience an emergent cosmology which
is a reflection of the boundary cosmology.
The general expression for the effective energy density $\rho$
on the RSII brane is rather complicated but simplifies considerably for
$z_{\rm br} =\ell$. In this case
\begin{equation}
\frac{\rho}{\sigma_0}=\left|\frac{1+\rho_{\rm h}/\sigma_0-
 \epsilon \sqrt{1-2\rho_{\rm h}/\sigma_0-\mu\ell^4/a_{\rm h}^4}}{1-\rho_{\rm h}/\sigma_0-
 \epsilon \sqrt{1-2\rho_{\rm h}/\sigma_0-\mu\ell^4/a_{\rm h}^4}}\right|
 -\frac{\sigma}{\sigma_0} ,
 \label{eq3330}
\end{equation}
where $\epsilon$ may take the values $+1$ or $-1$ .
Hence, the function $\rho=\rho(\rho_{\rm h},a_{\rm h})$ is not uniquely defined, although the mapping
$a_{\rm h}\rightarrow a$ is unique (Fig.~\ref{fig3})

For an arbitrary $z_{\rm br} \neq\ell$ in the low density regime (relevant for the one
sided version only), i.e., $\rho_{\rm h}^2 \ll \sigma_0^2$ and $\mu \ell^4 \ll a_{\rm h}^4$ we find: \\
a) For $\epsilon=-1$  at linear order in $\mu$ and quadratic order in $\rho_{\rm h}$
the effective energy density 
\begin{eqnarray}
 \frac{\rho}{\sigma_0} &=& 1- \frac{\sigma}{\sigma_0}
 + \frac{z_{\rm br}^2}{\ell^2}\frac{\rho_{\rm h}}{\sigma_0}
 +\frac12  \frac{z_{\rm br}^2}{\ell^2}\left(\frac{z_{\rm br}^2}{\ell^2}+1\right)\frac{\rho_{\rm h}^2}{\sigma_0^2}
 \nonumber
 \\
 &&
 -\frac12  \frac{z_{\rm br}^2}{\ell^2}\left(\frac{z_{\rm br}^2}{\ell^2}-1\right)\frac{\mu \ell^4}{a_{\rm h}^4}  + \dots .
 \label{eq3317}
\end{eqnarray}
and
 pressure 
\begin{equation}
p=-(\sigma_0-\sigma) +\frac{z_{\rm br}^2}{\ell^2} p_{\rm h}+\dots .
 \label{eq3320}
\end{equation}
Hence, at linear order the effective fluid on the RSII brane satisfies the
same equation of state as the fluid on the holographic brane. The
cosmological constant term will vanish on both branes if the RSII fine
tuning condition is imposed.
\\
b) For $\epsilon=+1$ at linear order
  \begin{equation}
 \frac{\rho}{\sigma_0}=\frac{z_{\rm br}^2/\ell^2 +1}{z_{\rm br}^2/\ell^2 -1}-\frac{\sigma}{\sigma_0}
 +\frac{z_{\rm br}^2/\ell^2}{(z_{\rm br}^2/\ell^2 -1)^2}
  \frac{\rho_{\rm h}}{\sigma_0}
- \frac{z_{\rm br}^2/\ell^2}{2(z_{\rm br}^2/\ell^2 -1)^3} \frac{\mu \ell^4}{a_{\rm h}^4} + \dots .
 \label{eq3324}
\end{equation}
Hence, in this case, the effective energy density $\rho$ on the RSII brane differs from $\rho_{\rm h}$ on the
holographic brane by a multiplicative constant and diverges in the limit $z_{\rm br}\rightarrow \ell$.
The effective cosmological constant on the RSII brane does not
vanish even if $\sigma=\sigma_0$ in which case 
\begin{equation}
\Lambda_{\rm br}=\frac{6}{\ell^2}\frac{z_{\rm br}^2/\ell^2 +1}{z_{\rm br}^2/\ell^2 -1}-\frac{6}{\ell^2} .
 \label{eq3312}
\end{equation}

\section{Conclusions \& Outlook}

Our study can be summarized as follows:
\begin{itemize}
\item
We have explicitly constructed the mapping between two
cosmological braneworlds: holographic and RSII .
\item
The cosmologies are governed by the corresponding modified
Friedmann equations.
\item
There is a clear distinction between 1-sided and 2-sided
holographic map with respective 1-sided and 2-sided versions of
RSII model.
\item
In the 2-sided map the low-density regime on the two-sided
RSII brane corresponds to the high negative energy density on
the holographic brane.
\item
The low density regime is maintained on both branes only in
the one-sided RSII
\item
We have  analyzed the effective energy density in two scenarios:
the RSII scenario with
the primary braneworld  at $z=z_{\rm br}$ and
the holographic scenario with 
the primary cosmology  on the AdS
boundary at $z = 0$.
Then, in the holographic and RSII scenarios we will have 
emergent cosmologies on the RSII and holographic branes, respectively.
\end{itemize}

It is conceivable that we live in a braneworld with emergent
cosmology. That is, dark energy and dark matter could be
emergent phenomena induced by what happens on the primary
braneworld.
In this regards, the holographic scenario offers a few interesting possibilities.
For example, suppose our universe is a one-sided RSII
braneworld the cosmology of which is emergent in parallel
with the primary holographic cosmology.
If the energy density $\rho_{\rm h}$ on the holographic brane describes matter with the equation of state
satisfying $3p_{\rm h}+\rho_{\rm h}>0$, as for,  e.g., cold dark matter,
according to (\ref{eq3324}) and (\ref{eq3312}) we will have 
an asymptotically de Sitter universe on the RSII brane.
If we choose the curvature radius $\ell$ so that the cosmological constant 
$\Lambda_{\rm br}$ fits the observed value, the
quadratic term will be comparable with the linear term today but
will strongly dominate in the past and hence will spoil the
standard cosmology. However, the standard $\Lambda$CDM cosmology
could be recovered by including a negative cosmological
constant term in $\rho_{\rm h}$ and $p_{\rm h}$ and fine tune it to cancel $\Lambda_{\rm br}$ up to a small
phenomenologically acceptable contribution.

\section*{Acknowledgments}

This work has been supported by the Croatian Science Foundation under the project 
IP-2014-09-9582 and partially supported by the H2020 CSA Twinning project No.\ 692194, ``RBI-T-WINNING''
and by ICTP - SEENET-MTP project NT-03 
Cosmology - Classical and Quantum Challenges.

\end{document}